\documentclass[Journal]{IEEEtran}
\usepackage{cite}
\usepackage{epsfig,graphics,mathrsfs,amssymb,amsmath,bm,color,yfonts,txfonts,multirow,multicol,slashbox,dblfloatfix}
\usepackage{subfigure}
\usepackage{algorithm2e}
\usepackage[noend]{algpseudocode}
\usepackage{flushend}

\begin{document}

\title{Big Data Analytics, Machine Learning and Artificial Intelligence in Next-Generation Wireless Networks }
\author{Mirza Golam Kibria, Kien Nguyen, Gabriel Porto Villardi, Ou Zhao, Kentaro Ishizu and Fumihide Kojima
 \thanks{The author is with Wireless Systems Laboratory, Wireless Networks Research Center, National Institute of Information and Communications Technology (NICT), 3-4 Hikarino-oka, Yokosuka 239-0847, Japan 
  (e-mails: $\text{\{mirza.kibria, kienng, gpvillardi,  zhaoou, ishidu, f-kojima\}@nict.go.jp}$). }
}
\maketitle

\begin{abstract}
The next-generation wireless networks are evolving into very complex systems because of the very diversified service requirements, heterogeneity in applications, devices, and networks. The mobile network operators (MNOs) need to make the best use of the available resources, for example, power, spectrum, as well as infrastructures. Traditional networking approaches, i.e., reactive, centrally-managed, one-size-fits-all approaches and conventional data analysis tools that have limited capability (space and time) are not competent anymore and cannot satisfy and serve that future complex networks in terms of operation and optimization in a cost-effective way. A novel paradigm of proactive, self-aware, self-adaptive and predictive networking is much needed. The MNOs have access to large amounts of data, especially from the network and the subscribers. Systematic exploitation of the big data greatly helps in making the network smart, intelligent and facilitates cost-effective operation and optimization. In view of this, we consider a data-driven next-generation wireless network model, where the MNOs employ advanced data analytics for their networks. We discuss the data sources and strong drivers for the adoption of the data analytics and the role of machine learning, artificial intelligence in making the network intelligent in terms of being self-aware, self-adaptive, proactive and prescriptive. A set of network design and optimization schemes are presented with respect to data analytics. The paper is concluded with a discussion of challenges and benefits of adopting big data analytics and artificial intelligence in the next-generation communication system.

\end{abstract}

\IEEEpeerreviewmaketitle
%==========================================================================
\section{Introduction}

In a service-driven next-generation network, a single infrastructure needs to efficiently and flexibly provide diversified services such as enhanced mobile broadband, ultra-reliable and low-latency communications and massive machine type communications. It should also support coexistent accesses of multiple standards such the fifth generation (5G), long-term evolution (LTE) and Wi-Fi, and coordinate a heterogeneous network with different types of base stations (BSs), for example, macro, micro, femto, pico BSs and diverse user devices as well as applications\cite{Boccardi}. The challenge to efficiently operate a network capable of facilitating such flexibility while satisfying the demands from diversified services is huge for a mobile network operator (MNO). On top of this, the MNOs face huge challenges in extending their coverages and keeping up with the ever-increasing capacity demands with a limited pool of capital and scarcity of resources such as spectrum.  Manual configuration for network planning, control, and optimization will make things even more complex. Moreover, the human-machine interaction can, sometimes, be time-consuming, susceptible to error and expensive. Consequently, automation of various entities and functions of the cellular networks has been one of the principal concerns of the MNOs in consideration of reducing the operational expenses.

\begin{figure*}
  \centering
   \includegraphics[scale=.523]{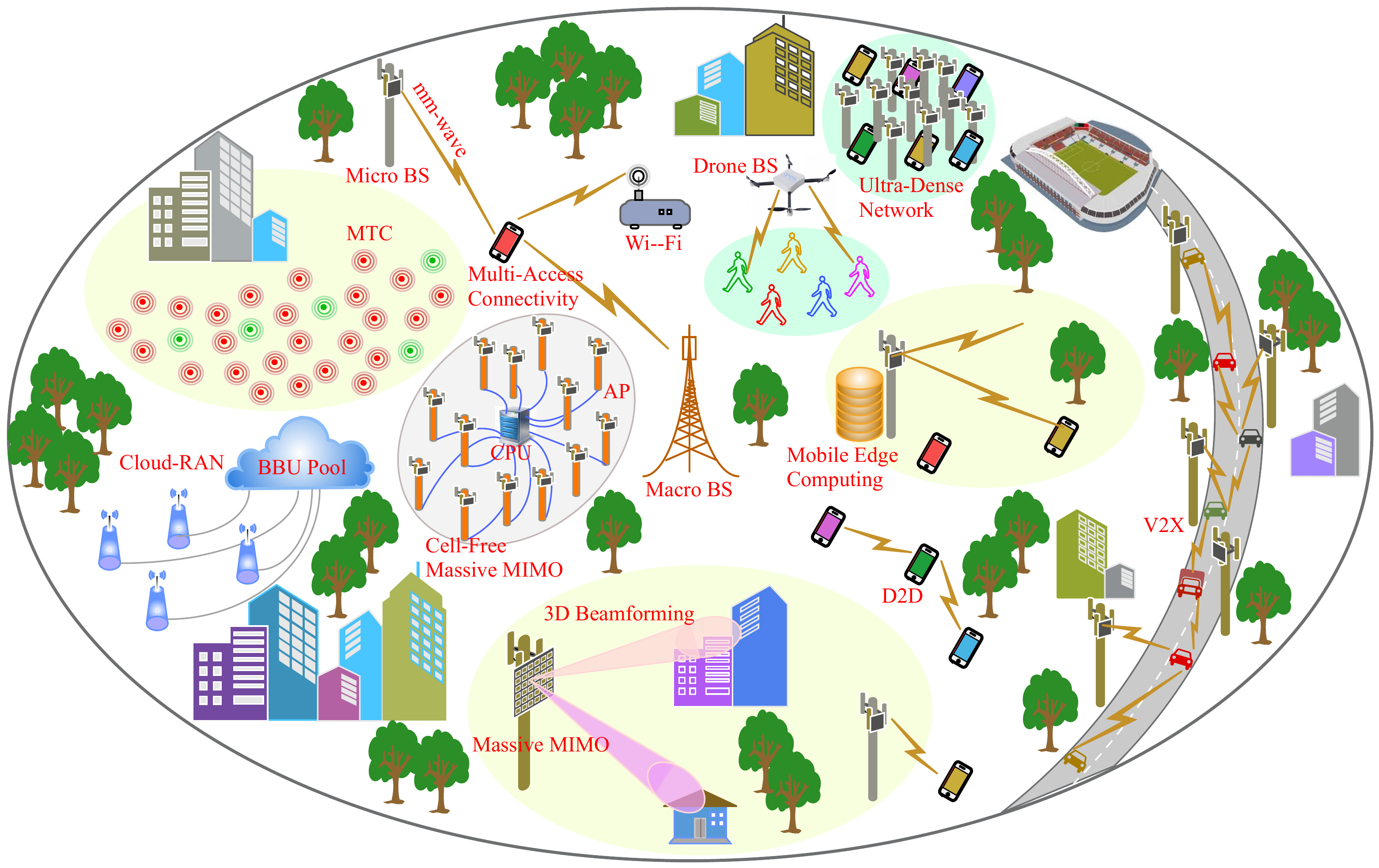}
   \caption{Graphical illustration of the next-generation communication system with some technological elements. Slicing, virtualization, (mobile) edge-computing, massive multiple-input multiple-output (MIMO), 3-dimensional (3D) beamforming, (ultra-dense) small cell networks, device-to-device(D2D) cell-free massive MIMO, multi-connectivity, cloud-radio access network (RAN), millimetre-wave, cloud-architecture/computing, etc., are the fundamental technologies to achieve the targets of the next-generation network.}
   \label{fig_Com_0}
\end{figure*}

Operators have been optimizing their networks all along, but even today, the prevailing approach is to independently optimize single key performance indicators (KPIs), or an element within the network independently\cite{Paolini}, thus using a small number of data sources. The MNOs mostly depend on KPIs accumulated at different locations/parts of the network to make decisions employing various data analysis tools.
Network monitoring and optimization are still predominantly performed on old/recorded data, but this greatly restricts their capacity. The MNOs, in general, have/can have access to a huge amount of data from their own networks and subscribers. With the appropriate analytics, big data can convey broader intuitiveness and understanding since it draws from multiple sources to reveal previously unknown patterns and correlations\cite{Bi}. It benefits to acquire a thorough understanding of various unknown values and delivers new measures in enhancing the performance from different levels of wireless networks. 

The value that analytics brings to optimization comes from expanding the range of data sources and taking a customer-centric, quality of experience (QoE)-based approach to optimizing end-to-end network performance. In widening the range of data sources, analytics requires more effort than traditional optimization, but it also provides a unified and converged platform for multiple targets of optimization. Now, within the 3rd generation partnership project (3GPP), network data analytics (NWDA) has been introduced to deliver slice and traffic steering and splitting (between 3GPP and non-3GPP access) related analytics automatically\cite{3GPP}. The european telecommunications standards institute (ETSI) has created the industry specification group called experimental network intelligence (ENI) that defines a cognitive network management architecture based on artificial intelligence (AI) techniques and context-aware policies. The ENI model helps the MNOs in automating the network configuration and monitoring process. 

From the operational expenses point of view, the system needs to be smart, self-aware, self-adaptive and must be able to run the network services economically and manage and operate the networks autonomously\cite{Han}. Conventional reactive maintenance is no more efficient. With big data analytics, the predictive and proactive maintenance of the network elements can be performed. With the volume of the data, the speed of data flowing in and the range and type of data sources, the network even go beyond prediction, i.e., it can assist and or prescribe the operations and maintenance unit with decision options and impacts of the actions, etc. Machine learning (ML) and AI can help in uncovering the unknown properties of wireless networks, identify correlations and anomalies that we cannot see by inspection, and suggest novel ways to optimize network deployments and operations.

\section{Drivers and Evolution of Analytics in Next-generation Wireless Systems and Computational Intelligence }

  \subsection{Drivers to Adoption of Analytics } 
  
 The ever-increasing complexity of the networks and complicated traffic patterns make the big data analytics appealing and very important for the MNOs. The MNOs were earlier very cautious about the adoption of big data analytics, however, multiple drivers are turning the MNOs cautious stance towards the comprehension that deep optimization of the networks and the services are extremely necessary for near future. As a consequence, there exists a consistent and rational commitment to capturing a deep knowledge and understanding of the network dynamics and make the best use of them through optimization. Three predominant drivers strengthening the adoption of big data analytics \cite{Paolini} can be identified as {\it Cost and Service} drivers, {\it Usage} drivers and {\it Technology} drivers. In the following, we discuss them in more details.

{\it Cost and Service drivers}: The subscribers, in general, are more demanding but less eager to raise the wireless pay out. In such environment, there is an urgent need for optimization of the usage of network resources. Furthermore, the network-centric service model is transforming into a user-centric service model based on the QoE. As a result, the MNOs need to better understand the QoE and its relationship with the network's KPIs. In addition, the MNOs need to retain its customers. As a result, the MNOs need to (i) manage its traffic based on service and application, (ii) improve efficiency to retain profit margins, (iii) improve network performance and QoE without increasing cost and (iv) keep churn as low as possible, etc.  

{\it Usage drivers}: The traffic patterns, subscriber equipments, and subscribers' profiles are all heterogeneous in nature. In a user-oriented service model, analytics supports the MNOs maintain and regulate traffic types, wireless devices, and subscribers diversely based on the MNOs' strategies and each's requirements. Furthermore, the wireless traffic load is growing faster than the capacity, and the MNOs are facing tough challenges to increase network capacity in a cost-effective way. Therefore, intensifying the resource utilization is required. Analytics take the network load into account and helps the MNOs to manage network traffics more efficiently in real time.

{\it Technology drivers}: The next-generation wireless networks have many technology components such as network resource virtualization, edge-computing, mobile edge-computing, network-slicing, etc. It integrates multiple air-interfaces, network layers and accommodates a range of use-cases. The MNOs need robust analytics framework to orchestrate the virtualized network resources efficiently. The analytics also help the MNOs to balance the centralized and distributed functionality. The data analytics facilitates the MNOs to figure out the most competent way to slice the network and traffic, i.e., the number of slices, splitting traffic across slices, etc, which depend on the type of traffic and how varies over time and space. 
 \begin{figure}
  \centering
   \includegraphics[scale=.523]{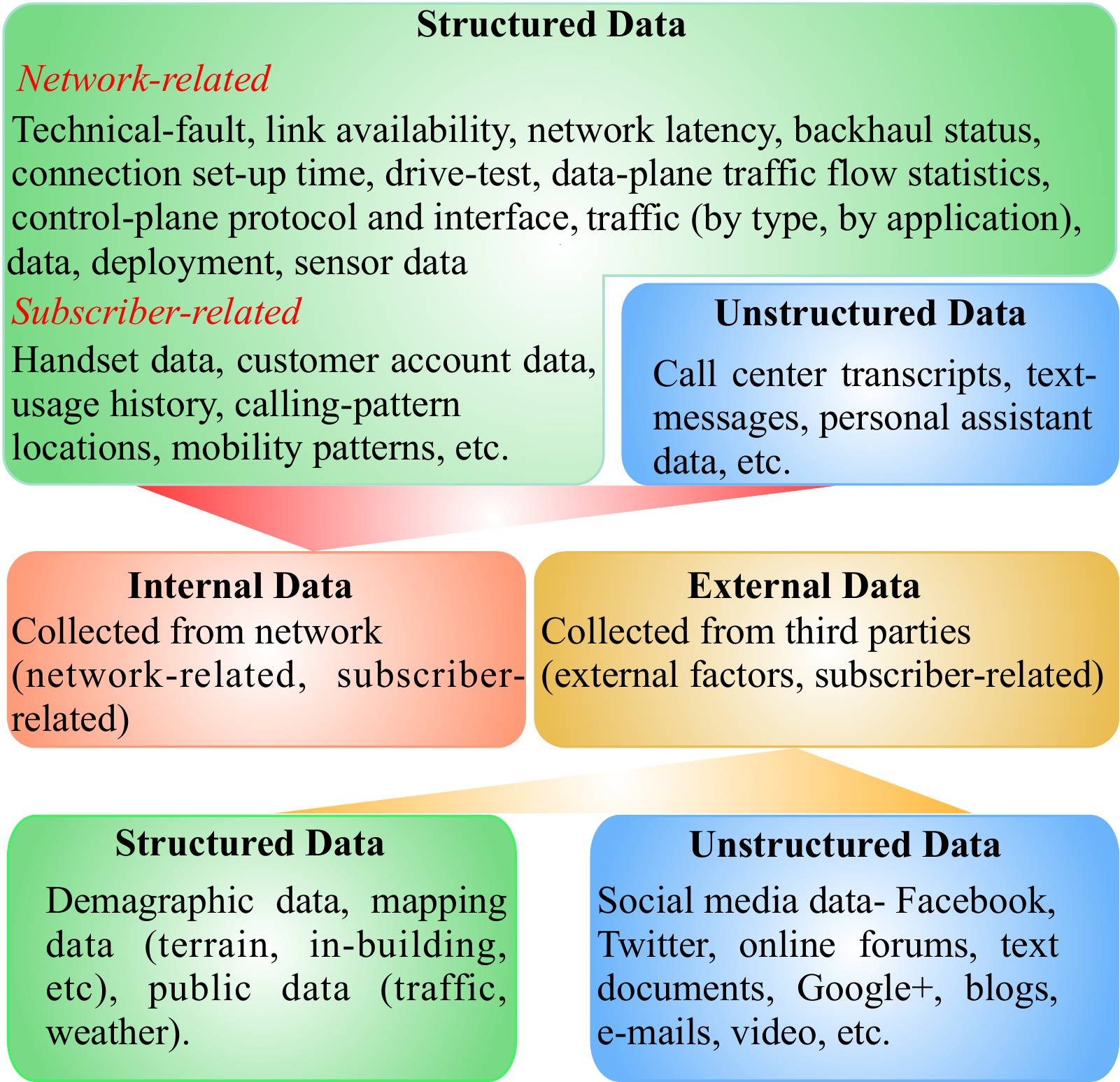}
   \caption{Data sets and the sources of data available to the MNOs for big data analytics, machine learning and artificial intelligence.}
   \label{fig_Com_1}
\end{figure}

\begin{figure*}
  \centering
   \includegraphics[scale=.55]{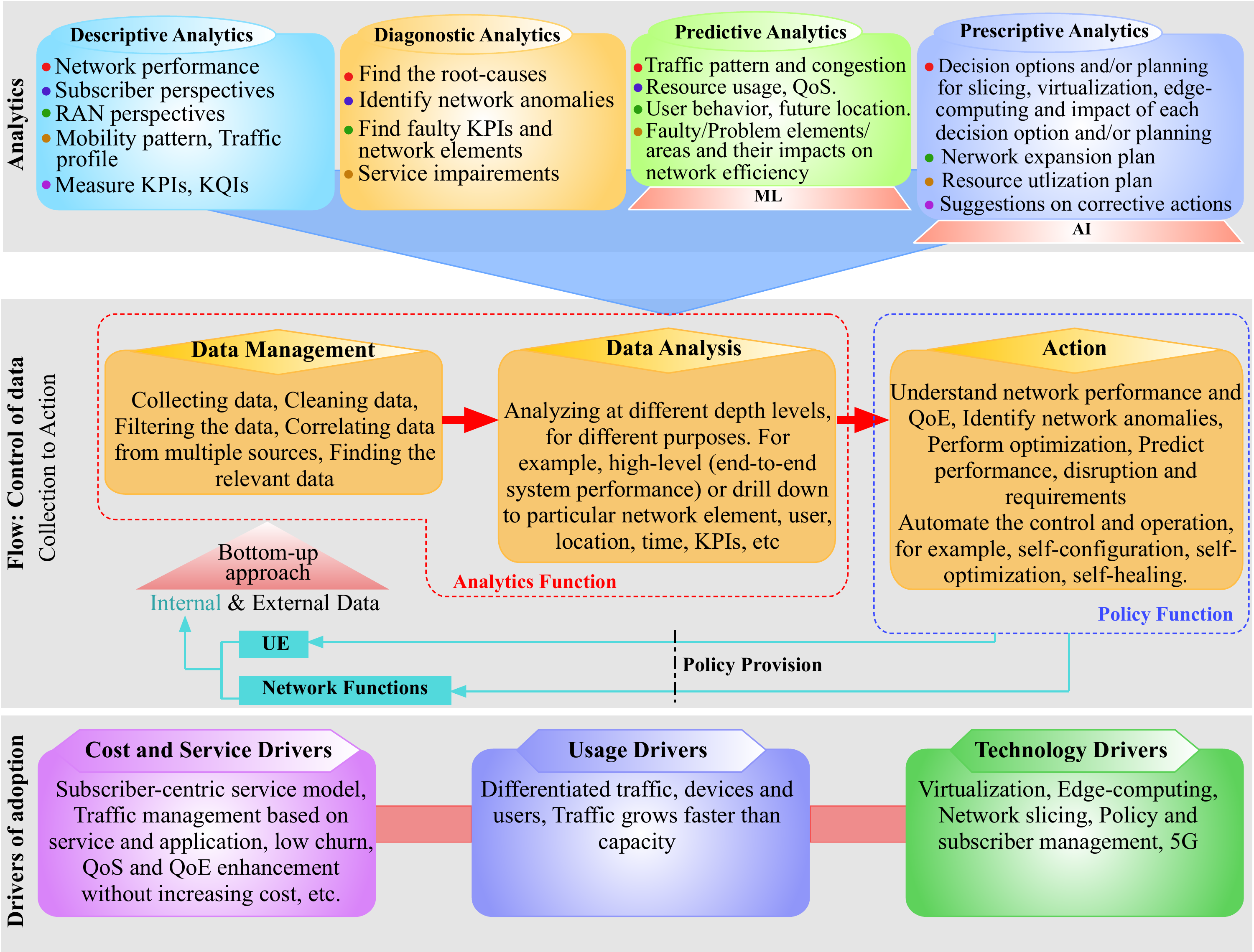}
   \caption{Drivers from different directions that strengthen the case for the adoption of analytics in next-generation communication systems. Flow of processes/actions of the MNO that employs analytics. Here, the Analytics Function collects the data and analyze them. The Policy Function obtains the analytics reports produced by the Analytics Function and may dynamically and intelligently deliver analytics-based policy rules for UE and the network functions.}
   \label{fig_Com_2}
\end{figure*}

\subsection{Types of Analytics}
There exists a succession of evolution in big data analytics, starting from descriptive analytics to diagnostic analytics to predictive analytics, and excelling towards prescriptive analytics as shown in Fig.~\ref{fig_Com_2}, out of which three (descriptive, predictive and prescriptive analytics) are dominant. The MNOs currently are in descriptive phase and use mainly the visualization tools to get insights on what has happened, the network performance, traffic profile, etc. The MNOs can make use of the diagnostic analytics to figure out the root-causes of the network anomalies and find out the faulty KPIs and network functions/elements. In order to get the diagnostic analytics, the analytics tool employs techniques like drill-down, deep learning, data discovery, correlations, etc. 

Predictive analytics is a great tool for making predictions. Note that it can never report or be certain about what will happen, however, predictive analytics can only produce forecasting about what might happen, for example, future locations of the subscribers, future traffic pattern and network congestion, etc. Predictive analytics deliver predictions about the future events based on the real-time and archived data by making use of various statistical techniques such as machine learning, data mining, modeling as some statistical process and game-theoretic analysis. Prescriptive analytics goes steps ahead of just predicting the future events by suggesting decision options for slicing (i.e., how to slice, how many slices), virtualization, edge-computing, etc., along with the implications of each decision option. Therefore, the prescriptive analytics need an efficient predictive model, actionable data and a feedback system for tracking down the results generated by the action taken. The decision options(e.g., for network expansion, resource usage) are produced considering the MNOs preferences, system constraints (backhaul, fronthaul, spectrum, transmission power), etc. Prescriptive analytics can also suggest the finest course of actions for any pre-defined target, for example, of a particular KPI.

The MNOs have access to large amounts of data which can be categorized into two classes such as internal data and external data as shown in Fig.~\ref{fig_Com_1}. The internal data corresponds to data belonging to the MNOs and/or produced in the network, which is network related and subscriber related. The external data is collected from the third parties. Both the internal and external data can be further classified into two categories, which are structured data and unstructured data. The structured is stored in a relational database, i.e., each field in the database has a name and the relationship between the fields are well-defined. On the other hand, the unstructured data (for example, call center transcripts, messages, etc) is not usually saved in a relational database. A comprehensive coverage on the features and sources of mobile big data can be found in \cite{Cheng1, Cheng2}.

\subsection{Computational Intelligence}
MNOs have access to a collection of data sets (i.e., these data can be highly dimensional, heterogeneous,
complex, unstructured and unpredictable) that are so large and complex that the traditional data processing and analysis approaches cannot be employed due to their limited processing space and/or processing time. Computational intelligence, a set of nature-influenced computational techniques and approaches, play a very crucial role in the big data analysis\cite{Engelbrecht}. It enables the analytics agent to computationally process and analyze the historical and real-time data, and eventually finds out and explain the underlying patterns, correlations, as well as to intensely understand the specific tasks. The computational analysis tools and methodologies convert the MNOs' massive amount of raw data (unprocessed, structured/unstructured) into meaningful data/information.

For feature selection, data-size and feature space compliance, active-incremental-manifold-imbalance learning on big data, uncertainty modeling, sample selection, classification/clustering, etc., there are many tools and methodologies that can be applied for big data analysis. For example,  
fuzzy logic, neural algorithms, rough sets, swarm intelligence, evolutionary computing, stochastic algorithms, physical algorithms, immune algorithms, learning theory, probabilistic methods are the tool and methodologies that the MNOs' big data analytics agents can employ for computationally processing and analyzing the available data.

In general, for big data analytics, the MNO can follow two distinct approaches, namely top-down approach and bottom-up approach\cite{Acker}. In the top-down approach, the MNOs define their targets to be achieved or problems to be resolved, and then decide what data sets are required. Whereas, in the bottom-up approach, the MNOs already have access to massive amounts of data and then exploit the big data on hand to get the insights. The top-down approach delivers incremental benefits and it is very challenging to execute. It also, in most of the cases, does not bring on surprising and adventitious results. On the other hand, the bottom-up approach facilitates a more outright and transparent view of the network performance, subscribers' behaviors, resource utilization, etc., and may bring on completely new opportunities for the MNOs. The bottom-up approach is also likely to capture the subscribers' perspectives the RAN perspectives and may beget new business opportunities for the MNOs.

 \section{Machine Learning and Artificial Intelligence for Managing Complexity} 
  
The ML and AI are two very powerful tools that are emerging as solutions for managing large amounts of data, especially for making predictions and providing suggestions based on the data sets. They are, however, very often appear to be used interchangeably in spite of some parallels. ML is sometimes brought up as a subspace of AI based around the concept that we can let the machines learn for themselves by providing them access to large amounts of data. On the other hand, AI is the extended and wider perception of machines becoming capable of carrying out tasks in an intelligent way. Compared to the generalized AI (a generalized AI system, in theory, can handle any task), applied AI is more suitable for next-generation communication systems as the applied AI system can be devised to adeptly controlling and optimizing the wireless networks. Unlike ML models, AI models reach out the world, accustom to the changes and rebuild themselves\cite{Paolini}. While ML is great for predictive analytics\cite{Jiang}, AI goes beyond predictions and prescribe plans/suggestions with implications to realize a benefit.

Managing wireless networks that grow in size and complexity becomes very difficult since there is need to integrate new elements and technologies to benefit from the technological advances. The amount of data such large and complex networks produces are too large and too complex. Machine learning and artificial intelligence are useful for analytics as they can extract valuable information from the raw data and generate insightful advice and predictions. ML and AI are expected to assume the primary role in the development and evolution of analytics, but analytics will not reduce to them. ML is largely developed from AI, hence the two overlap. ML has tools to extract relevant information, suggestions, and predictions from the data sets that are too large and too complex. While AI has a wider scope: to replicate human intelligence or some aspects of it and other cognitive functions. 

Furthermore, for non-recurring events, there is no historical data to rely on, hence the real behavior of the network will diverge from the predictions\cite{Paolini}. The ML and AI are becoming potential to help MNOs to address areas which are new and there are no historical data, or too complex to understand with traditional approaches. The ML and AI tools can correlate multiple sources of data and find what is relevant. They may also reveal interrelations and dependencies that were not previously identified because their automated mechanisms have the capability of anatomizing and inspecting data more intensely and more methodically. Although human expertise is useful in confining the focus to produce solutions and to manage complex problems, it has limited capability in finding novel solutions and insights. The future of wireless networks will undoubtedly rely on AI. In \cite{Chen}, the authors have provided a panned overview on the range of wireless communication problems and issues that can be efficiently addressed using AI while providing detailed examples for the use-case scenarios.

\section{Data-Driven Coverage and Capacity Optimization of Next-Generation Cellular Wireless Networks}
The conventional network-centric architecture cannot capture all of the nuances that can affect service quality. Mobile operators need solutions that provide them with an analysis capability that captures all the information relating to the network and subscribers into a single enterprise geolocation platform that can help remove the assumptions involved in fault isolation and reduce mean time to repair. The MNOs are suitably positioned to exploit big data analytics because of their access to huge amounts of data. The big data analytics engine/agent can produce/predict the following analytics based on its data, primarily from two sources, such as the network data and the subscriber data, which are then exploited to design and optimize the network.

\begin{figure*}
  \centering
   \includegraphics[scale=.53]{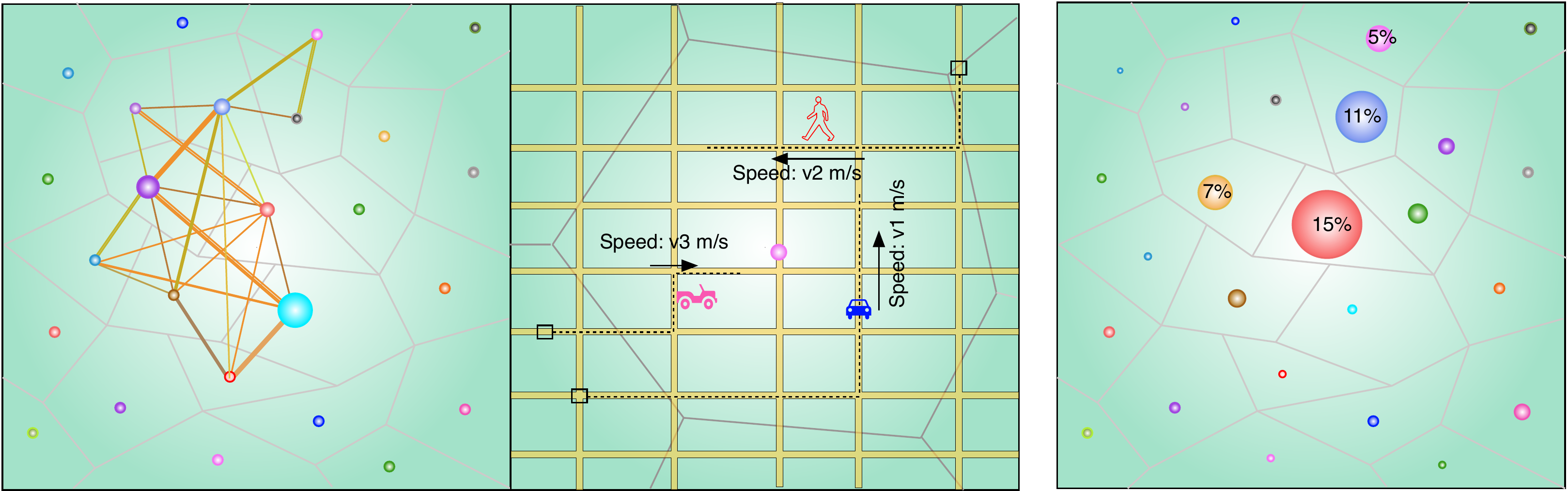}
   \caption{Left: Example of the trajectory of a mobile user who visited the vicinity of during an observational period. The circles correspond to the serving BSs, i.e., give the approximate locations. The gray lines depict the Voronoi lattice. The width of the line between two BSs is proportional to how often the user moves between the BSs and the sizes of the circles are proportional to the resources/traffic consumed. Middle: Example of the small-scale (e.g., within few seconds/minutes) trajectory of the user for more accurate/exact location of the user. Right: The circles represent the BSs and sizes of the circles are proportional to the traffic loads on the BSs, i.e., the traffic pattern and/or congestion status in the network can be measured. Note that results here do not comes from a real field experience.}
   \label{fig_HCSN_3}
\end{figure*}

\begin{itemize}

\item {\bf Subscriber Profile (SP)}: 
In this context, the subscriber profile consists of the device profile, service-level agreement (SLA), subscriber's affordability (price per unit of data-rate), quality of service (QoS)/policy, behavioral profile, etc. It plays a vital role in the abovementioned controlling and optimization process. The priority of the subscriber in the network is defined in the subscriber profile when resource allocation, congestion control and traffic offloading is performed. 
Behavioral data provides information how the user behaves in using various applications/services. For example, how frequently and when the user makes video/audio call and the average length of the call duration? Through analytics, we can speculate on a lot of these user attributes.

\item {\bf Subscriber Perspective (Sub-P)}: Subscriber perspective is an attribute/measure that associates MNO's network activity with the user's SLA, pricing, QoS, QoE, etc., and it delivers a subscriber-centric outlook of the network for analytics\cite{Kyriazakos}. Subscriber perspective is, in general, defined by the {\it Cost Over Quality Ratio}, which sometimes gets polished through a variety of attributes linked to the requested service class and perceived friendliness to the service, i.e., QoS violation, delay violation, etc.  It enables the MNO to measure or make a perception about the RAN quality from the subscriber point of view, and put them in a better position to provide a high QoE. 

\item {\bf RAN Perspective (RAN-P)}:  RAN perspective is a measure that provides the MNO the subscriber-centric RAN quality, i.e., the RAN performance from the subscriber's point of view\cite{Procera}. The user equipment's view of the signaling information such as signal strength, error codes, available networks, etc., are extremely helpful to the MNO for analytics. From the user's predicted trajectory, spatial deployment of the BSs and signaling metrics, the MNO can generate heat map for coverage and determine the RAN quality. Advanced cell mining that statistically analyzes the performance data enables the MNO to identify radio cell irregularities and other negative syndromes via anomaly (i.e., SLA violation) diagnosis and trend study of the time series data, and control traffic and RAN congestion problems. With RAN Perspective, the full end-to-end subscriber experience can be measured in terms of {\it Service Availability} and correspondingly mapped to the exact location in the network. The MNO can also use {\it Subscriber Satisfaction Coefficient} to define the RAN perspective. Note that the signaling metric cannot be easily retrieved from mobile gateways or retrieved by network probes. An efficient retrieving method is discussed in\cite{Procera} that uses SIM-based applet stored in users devices to collect the signal strength and quality metrics, thus the subscriber devices act as network probes in measuring the RAN perspective. 
\item {\bf Subscriber Mobility Pattern (SMP)}: In order to guarantee the QoS requirements and to efficiently maintain resources utilization, traffic offloading and routing, knowing the mobility information of a user in advance is very crucial.
Human travel pattern analyses reveal that people travel along specific paths with reasonably high predictability \cite{Lu}. The trajectory of a mobile user can be predicted based on user's present location, the movement direction and the aggregate history of SMP. It is possible to predict the spatiotemporal trajectory (trajectory with both spatial and temporal information), i.e., not only the mobile user's future location but also the time of arrival and the duration of stay can be predicted. Mobility pattern is generally based on user positioning, which can be estimated using the signals from the cellular system.

\item {\bf Radio Environment Map (REM)}: 
The MNOs can better plan, build, control and optimize their networks conforming to the spatiotemporal radio atmosphere, through prediction of radio signal attenuation. Many schemes have been developed that give the MNOs the means to predict the distribution of radio signal attenuation at different operating frequencies and in many different radio environments. The radio map along with the mobile user's predicted trajectory facilitates the prediction of average channel gains. There are several different methods to construct the radio map, for example, radio map based on drive test measurements, radio map based on measurements through user terminal equipped with global positioning system\cite{Atawia}. 

\item {\bf Traffic Profile (TP)}:  
In order to attain as well as predict the network's congestion status, tempo-spatial traffic load variation needs to be known, i.e., the knowledge of temporal traffic trace, BS spatial deployment and BSs' operating characteristics (transmission power, height, etc) are very important. The authors in \cite{Oh} report that the network's traffic load dynamics demonstrates periodical characteristics over days and hours, thus implying high predictability of the traffic load. The traffic profile along with the SMP can be used to estimate and predict the traffic arrival rate, congestion status of the network with required time resolution/granularity.

\end{itemize}

\begin{figure}
  \centering
   \includegraphics[scale=.45]{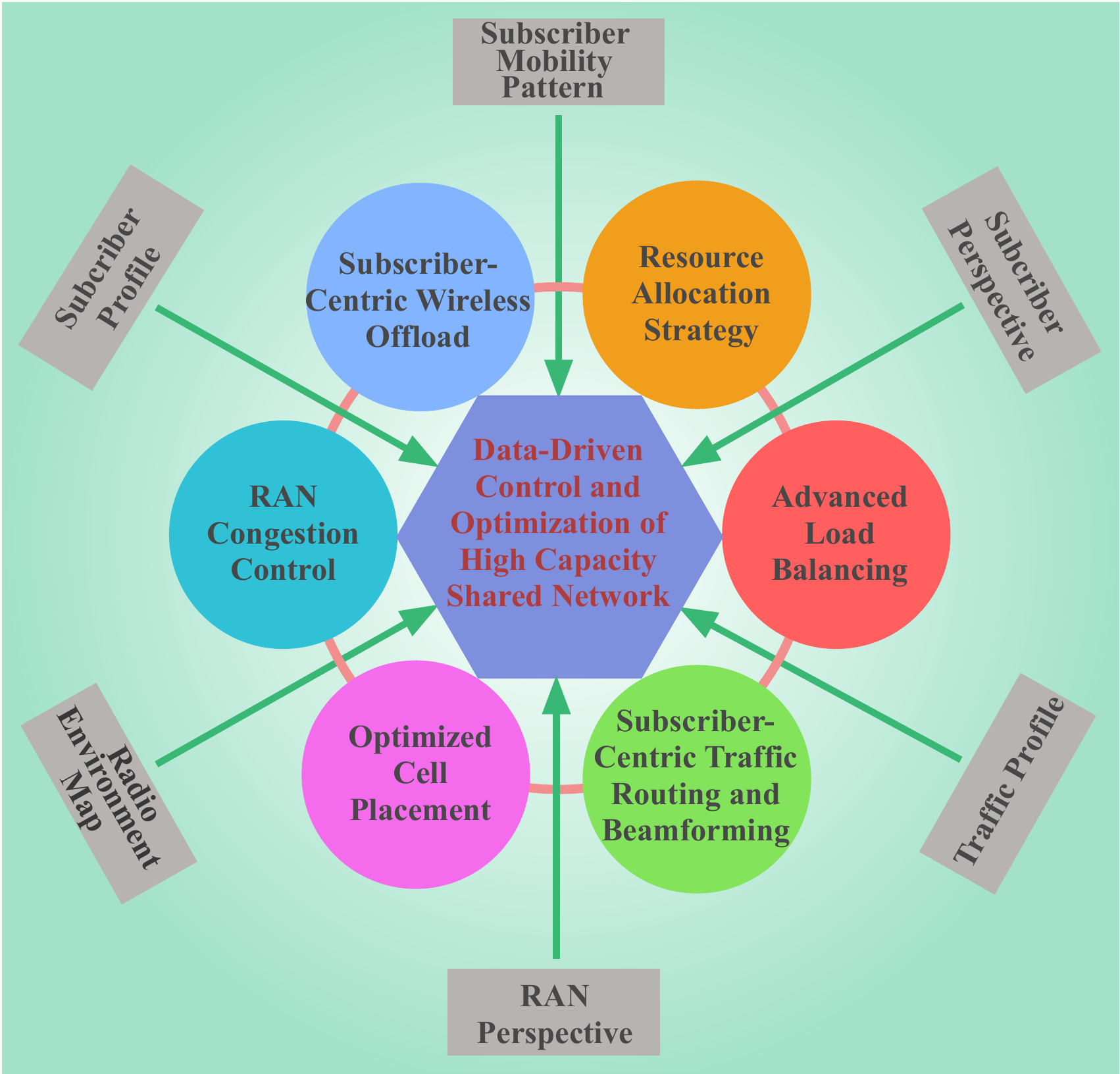}
   \caption{Some data analytics and their application in control and optimization of next-generation wireless communication systems.}
   \label{fig_HCSN_2}
\end{figure}

It is very crucial to have a sturdy, well-balanced load-distributed cellular system in a dynamic network and radio environment with mobile users using bursty applications and services. The next-generation network can employ the systems analytics, user and service analytics, radio analytics for control and optimization of the network\cite{Heavy} in the following scenarios.
\begin{itemize}

\item {\bf Resource Allocation Strategy} 
Advanced resource allocation is very crucial for enhancing the spectrum and power utilization efficiency of the communication systems. Leveraging the big data analytics based prediction ability in optimizing the resource allocation has been reported to be very advantageous. With the help of SMP and TP, the MNO can approximate the typical resource usage per-cell per-user in the network since (i) the average channel gains are predictable from the trajectory of the user and (ii) based on content popularity, user's behavioral profile and currently running application, the preferred contents can be predicted even before the individual users put forward their service requests. As a result, with the help of big data predictive analytics, the operators are able to predict changes in the users' service demands and thus can manage and optimize the resource allocation in real time. Integrated backhaul and access in mmWave 

\item {\bf Subscriber-Centric Traffic Routing}
Providing the best QoE as the end users' subjective perception is one of the most important requirements. Service delay, jitter affect the mobile users' QoEs very badly. Data-driven solutions can deliver traffic to different users depending on their subscription profile, types of applications, and preferences. A QoE-aware network continuously adopts the changing environment in order to deliver acceptable QoE. The SMP, the network utilization profile, and TP can help the operator to devise efficient routing protocol while considering the backhaul load, the SLA and the corresponding cost. Depending on users preferences and interests, and currently running application, the system can proactively cache the popular content, and use the backhaul route that is closer to the local caching server.

\item {\bf Subscriber-Centric Wireless Offload}
Due to an exponential surge in mobile data traffic carrier over macro cell layer, the MNOs are more and more finding out approaches to optimize the traffic in the network while ensuring seamless connectivity and minimum guaranteed QoS to its subscribers. Traffic offloading from macro cell layer to small cell layer (specifically towards WiFi networks) is a great way to relieve congestion in macro layer and enhance the overall network throughput. Blindly offloading the mobile users may result in dissatisfaction of the subscribers of the higher tier and breaching SLAs. Therefore, it is necessary to devise efficient solutions that aid the MNOs to decide and offload mobile users to WiFi, based on user profile and network congestion conditions. Data-driven contextual intelligence originated from correlating the customer profile (types of application, spending pattern, SLA) with SMP, TP, and REM, can decide which customer should be offloaded, and even to which small cell/WiFi the customer needs to be offloaded.

\item {\bf Optimized Cell Placement}
Small cell placement plays a vital role in defining the capacity of a heterogeneous network. Strategic small cell placement is very important in areas where subscribers concentrate while taking care of coverage goals, radio frequency interference issues and its potential in relieving congestion from the macro layer by offloading traffic. Rapidly placing the small cells at the very best locations is a complex issue as the number of small cells is much larger. Traditional macro cell layer management tool and even the self-organized networking tool may not compensate for improper cell placement. However, the data-driven solution can efficiently administer the small cell placement issue exploiting the knowledge of long-term user density, traffic intensity. Data-driven solutions incorporating the long-term TP, REM can devise optimized dynamic small cell placement strategy that identifies key locations where small cells need to be deployed and/or re-arranged in order to enhance the network capacity, minimize interference and improve the traffic offloading capability. The 3D geolocator tool that uses predictive ``finger printing" algorithms to locate traffic hotspots can simplify the cell placement task.

\item {\bf Radio Access Network Congestion Control}
The combination of limited network resources and ever-growing demands result in unavoidable RAN congestion, which degrades users' quality of experience. Expansion of existing RAN provides a solution to this problem, but it is expensive. A flexible, as well as a cost-effective solution is to deploy a proactive policy control mechanism preventing deficiency of RAN resources. Smart congestion control solution considering location information, the load level of network elements and users' service level agreements is able to deliver perceptibility at particular sub-cell level and caters priority to some set of subscribers based on their tiers. Since the congestion events are short-lived (typically congestion occurs at busy times of the day) and users future locations are predictable, with the help of data-driven predictive analytics incorporating the correlation between SMP, radio map and traffic profile, advanced proactive RAN congestion control mechanism can be deployed where the occurrence of RAN congestion is predicted. RAN congestion controlling can be done in many ways, for example, by reducing the QoS for subscribers belonging to the lowest tier of users, rejecting new session establishment, terminating certain sessions.

\begin{figure*}
  \centering
   \includegraphics[scale=.60]{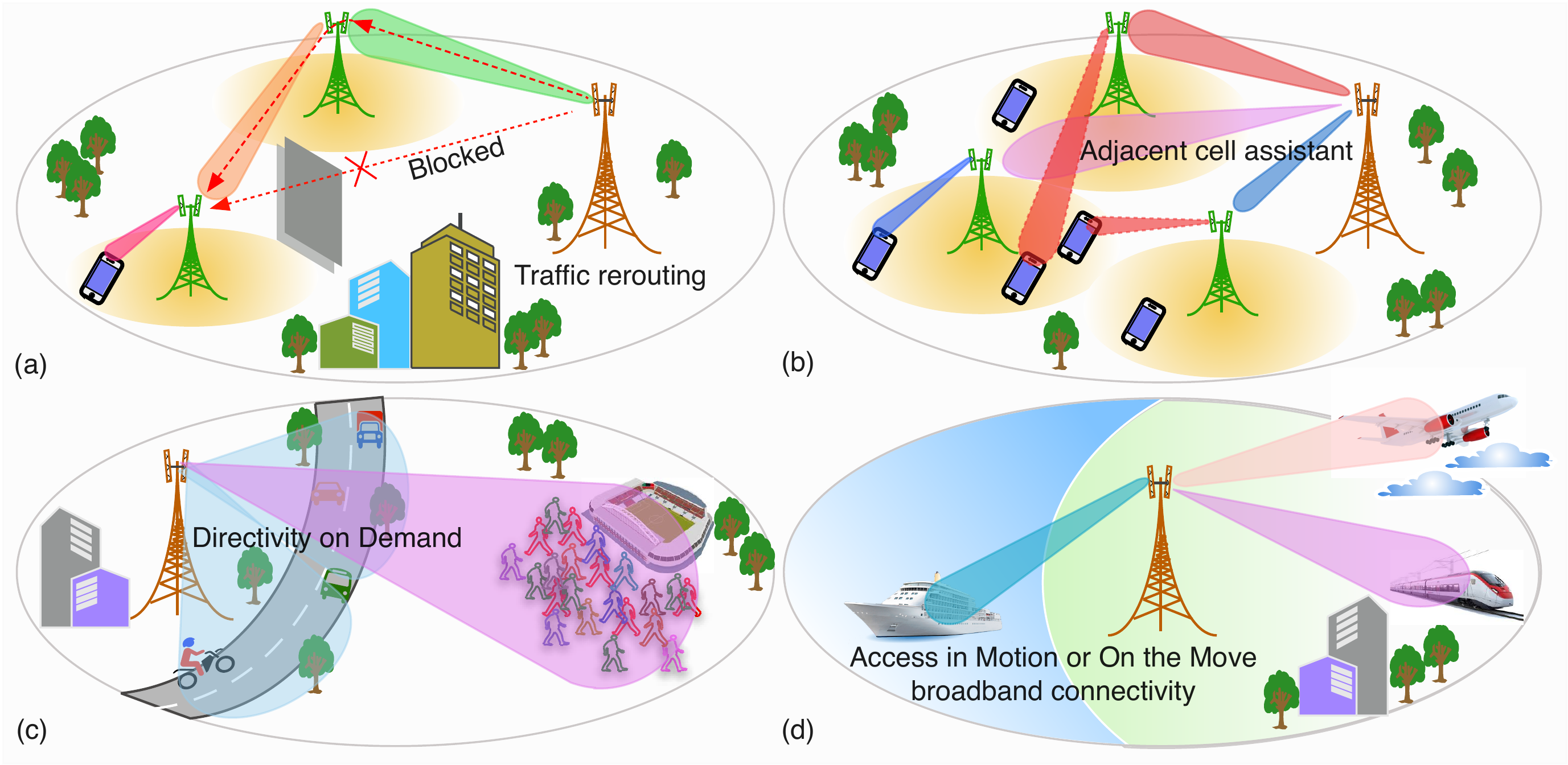}
   \caption{Data analytics and ML, and AI techniques can be used in analog, digital, and hybrid beamforming in terms of generating the optimal beam patterns, dynamically selecting the most suitable beam, and performing beam-steering operation. In this figure, holographic beamforming has been taken as the use-case scenario.   }
   \label{fig_Figure1}
\end{figure*}

\item {\bf Advanced Load Balancing}
Note that the profile of mobile users and traffic in each cell is distinctive, and the profiles change from time to time. When a number of users disassociate with one cell and move to the neighboring cell, the network's traffic load distribution, i.e., the traffic profile may change severely, and as a consequence, some cells in the network may get overloaded causing service downgrade. Currently, the load balancing methods employed by the MNOs are almost manual, thus not efficient, and at the same time, they are not accurate enough. Predictive analytics by data mining and correlating the network and/or subscriber data such TP and SMP can not only help in understanding the cells' current load situation, but also in identifying the heavily loaded parts of the network and predicting the traffic variation in advance. Consequently, the MNOs can perform advanced load balancing and cell planning by adding capacity, expanding the coverage of unloaded or lightly loaded cells in order to unburden the neighboring overburdened/overloaded cell. The data-driven advanced load balancing will enable the MNOs to optimize the utilization of available network resources.

\item {\bf Advanced Beamforming}
Beamforming is an integral technology component in next-generation communication systems for enhancing the coverage and data rates. 
A BS with multiple antennas can generate multiple beams simultaneously\cite{Hong}. Under static beamforming (fixed beam pattern without beam-steering), for a mobile user, the quality of the serving beam may deteriorate, and hence a different beam from the same BS (from same sector or different 
sector) or an adjacent BS that serves the user well needs to be selected. The ML can help the serving node to dynamically select the best beam for the user. The ML also enables dynamic switching ON/OFF the beams based on TP and SMP for energy and interference minimization. Holographic beamforming\footnote{In holographic beamforming, the complex propagating wave across surface scattering antenna or the transmitting aperture becomes a holographic profile, i.e., the collective profile across the antenna-array elements represents the desired hologram for transmission. } (with electronic speed beam-switching/beam-steering)\cite{Black} along with data analytics and ML can help in dynamically rerouting the traffic, dynamic adjacent cell access, steering coverage where it is needed to accommodate usage patterns, for example, rush hour traffic, events, etc., as shown in Fig.~\ref{fig_Figure1}. \\

Fig.~\ref{fig_Figure1}(a) contemplates adaptive/dynamic rerouting of traffic when there is an obstacle or physical interference between two communicating nodes. The rerouting path can be dynamically selected based on TP and other informations such as resource availability. Dynamic adjacent cell assistant in Fig.~\ref{fig_Figure1}(b) with holographic beamforming facilitates serving a distant user outside the general coverage area of the assisting BS when the original serving BS has bandwidth shortage/overloaded or becomes non-functional. Depending on the received informations and making use of TP and REM, the holographic beamforming antenna can dynamically configure a high directivity beam towards the distant user\cite{Black1}. Similarly, dynamically steering coverages where it is required as shown in Fig.~\ref{fig_Figure1}(c) combined with 3D geolocator tool and configuring long-range, high-capacity links along with electronic speed beam-switching to provide access in motion as shown in Fig.~\ref{fig_Figure1}(d) can be enhanced by analytics from internal data, external data and ML.

\end{itemize}

Apart from the control and optimization scenarios mentioned above, accurately and efficiently accomplishing the maintenance of the network elements, backhaul monitoring (potential bottlenecks in backhaul networks), fronthaul management and orchestration, intelligent network slicing, energy optimization, monitoring of critical network health variables are some of the key pain and troublesome issues the MNOs  are very often challenged with. With big data analytics, the predictive maintenance of the network elements can be performed. The predictive maintenance inspects the operational status of the network elements through sensors in real-time. With the help of big data analytics, the potential risks can be identified, thus the possible faults are found earlier. This helps the operation and maintenance team of the MNO to become proactive to work out predictive maintenance planning.

Furthermore, there are plenty of other applications of data analytics, ML, and AI in next-generation communication systems. For example, the MNOs can employ analytics for obtaining good insights about the physical layer\cite{Tim} and the medium access control (MAC) layer, for example, for the optimal constellations in interference channels where the optimal schemes are unknown, for the best beamformer, for pre-caching/buffering, for the most suitable forward error correction code, for the optimum MAC protocols, predictive scheduling, etc.
Intelligent wireless network architecture,  RAN optimization in terms of transmission control protocol (TCP) window optimization, mobility management optimization can also be achieved through the use of big data analytics\cite{IHan}. Data aided transmission, network optimization, for example, channel modelling, multiple user access and novel applications such as  unmanned aerial vehicle/drone communications, smart grid, etc., have been discussed in \cite{Lijun}.

 \section{Challenges and Benefits}
 
Although employing big data analytics for control and optimization of wireless networks is very attracting to the MNOs, it comes with some challenges. The process of managing and leveraging of a huge amount of data, designing algorithms for dynamic and effective processing of sizable data sets and then exploiting the insights from the data analytics in networks can pose unique challenges. The prime concerns for the MNOs emerge from the extent of effort, skills, and workforce needed to manage and operate a big data platform.  However, the most important and difficult challenge is more likely to stem from the loss of direct control that the MNOs still have over the wireless network. The loss of direct control is incurred from the combination of automation and real-time operations within the big data analytics framework. However, the huge complexity of the next-generation networks makes the automation inevitable, and handover or relinquish that level of direct control is imperative. On top of these, a substantial investment is necessary.

Despite all the challenges, the MNOs are more considerate towards data analytics platform since the challenges are outweighed by the benefits. The big data analytics infuses efficiency into the provisioning of services and end-to-end network. Analytics facilitates the MNOs to gain from the better planning, increased utilization of network resources, efficient maintenance of the network elements and lower operation costs. It gives the MNOs the flexibility to define and execute their own network utilization strategy. It helps the operators to make new service and offer plans that are suited to subscribers' needs. Although the MNOs are already performing these kinds of service provisioning, analytics delivers richer insights. Analytics helps to improve subscriber management and policy implementation. With the aid of natural language processing and interfacing with the smart digital assistants in the user devices, an autonomous customer care can be facilitated. Another benefit the MNOs get by employing analytics is differentiation, which is compelling to strengthening an MNO's market positioning. Analytics can support the MNOs to employ new techniques to traffic handlings such as network-slicing (i.e., the way to slice the network) and edge-computing (i.e., the way to balance centralized and distributed functionality).

\section{Conclusion}

We consider a data-driven next-generation wireless network model, where the MNOs employs advanced data analytics, ML and AI for efficient operation, control, and optimization. We present the main drivers of big data analytics adoption and discuss how ML, AI and computational intelligence play their important roles in data analytics for next-generation wireless networks. We present a set of network design and optimization schemes with respect to data analytics. Finally, we discuss the benefits and challenges that the MNOs encounter in adopting big data analytics.

\section*{Acknowledgement}
This research was conducted under a contract of R\&D for Expansion of Radio Wave Resources, organized by the Ministry of Internal Affairs and Communications, Japan.\\

The first author would like to thank Monica Paolini from Senza Fili Consulting for the insightful comments.

\end{document}